\documentclass[aps,amssymb,amsmath, twocolumn, pra, superscriptaddress]{revtex4}
\usepackage{graphicx}
\usepackage{epsfig}
\usepackage{dcolumn}
\usepackage{bm}
\usepackage{bbm}
\usepackage{hyperref}
\usepackage{color}
\usepackage[up]{subfigure}
\newcommand{\be}{\begin{equation}}
\newcommand{\ee}{\end{equation}}
\newcommand{\bc}{\begin{center}}
\newcommand{\ec}{\end{center}}
\newcommand{\bea}{\begin{eqnarray}}
\newcommand{\eea}{\end{eqnarray}}
\newcommand{\ba}{\begin{array}}
\newcommand{\ea}{\end{array}}

\begin{document}
\title{Two-state quantum walk on two- and three-dimensional lattices}
\author{C. M. \surname{Chandrashekar}}
 \email{cmadaiah@phys.ucc.ie}
\affiliation{Department of Physics, University College Cork, Cork, Republic of Ireland}
\begin{abstract}
We present a new scheme for a discrete-time quantum walk on two- and three-dimensional lattices using a two-state particle. We use different Pauli basis as translational eigestates for different axis and show that the coin operation, which is necessary for one-dimensional walk is not a necessary requirement for two- and three- dimensional walk but can serve as an additional resource. Using this scheme, the probability distribution from Grover walk using four-state particle and other equivalent schemes on a square lattice using coin operation is reproduced. We also present the Hamiltonian form of evolution  which can serve as a general framework to simulate, control, and study the dynamics in different physical systems.
\end{abstract}
\maketitle
\section{Introduction}
\label{intro}
Quantum walks, developed as the quantum analog of the classical random walks\,\cite{Ria58, Fey86, Par88, ADZ93} first emerged as a powerful tool in the development of quantum algorithms\,\cite{Amb03, CCD03, SKW03, AJR05}. Subsequently, its rich dynamics is constituting as a framework to understand and simulate the dynamics in various systems. For example, they have been used to explain phenomena such as the breakdown of an electric-field driven  system\,\cite{OKA05} and mechanism of wavelike energy  transfer  within photosynthetic systems\,\cite{ECR07, MRL08}, to demonstrate the coherent quantum control over atoms\,\cite{CL08} and localization of Bose-Einstein condensates in optical lattice\,\cite{Cha11a} and to explore topological phases\,\cite{KRB10}. The quantum walk evolution are widely studied  in two forms: continuous-time  quantum walk (CTQW)\,\cite{FG98} and discrete-time  quantum walk (DTQW)\,\cite{ADZ93, DM96, ABN01, NV01, BCG04, Ken07, CSL08}. In this letter we focus on the DTQW evolution which is defined on the {\it position} Hilbert space, ${\cal H}_p$ and the {\it coin} (particle) Hilbert space, ${\cal H}_c$. 
During last few years, experimental implementation of the DTQW has been demonstrated with energy levels in NMR\,\cite{RLB05}, 
ions\,\cite{SMS09, ZKG10}, photons\,\cite{PLP08, PLM10, SCP10, BFL10}, and atoms\,\cite{KFC09}. These experimental implementations using one-dimensional (1D) DTQW model on a two-level system using a degree two coin operation have opened up a new dimension to simulate quantum dynamics in physical systems like the recent demonstration of localization of photon's wavepacket\,\cite{SCP11}. Now the immediate interest would be to extend the implementation to two-dimensional (2D) and three-dimensional (3D) lattice structure with the available resources. This will give way to simulate and explore the possibility of mimicking the dynamics in various naturally occurring physical systems. One of the extension to the 2D is the Grover walk which is defined on a four-level particle with a specific initial state\,\cite{MBS02, TFM03}. An alternative extension to  higher ($d$) dimensions is to use a $d$ coupled qubits to describe the internal states\,\cite{ EMB05, OPD06}. This is extremely challenging with the available resources to implement it experimentally. To overcome this challenge, an alternative 2D DTQW using a two-state particle was very recently proposed. By evolving the particle in superposition of position space in one dimension followed by the evolution in the other direction using Hadamard coin operation  was show to be equivalent to four-state Grover walk\,\cite{FGB11}.
\par
In this letter we present a new scheme using different Pauli basis as translational eigenstate in different axis and show that the two-state particle walk can be implemented on a physically relevant  2D, square, triangular, kagome, and 3D, cubic lattice structures. Using basis vectors of the three Pauli matrices is very common in quantum optics experiments and various other physical systems making this scheme implementable in the present experimental setups. We also present Hamiltonian form of the evolution which can serve as a general framework to simulate, control, and study the dynamics in different 2D and 3D physical systems. We then show that the coin operation which is required for 1D DTQW is not a necessary requirement for 2D and 3D DTQW but can be used as an additional degree of freedom to control the dynamics. In particular, we demonstrate that the probability distribution obtained using Grover walk and alternative two-state walk on a square lattice \cite{FGB11} is effectively reproduced without a coin operation in our scheme. This further reduces the resource required for the experimental implementation.
\section{One-dimensional DTQW and its Hamiltonian form}
\label{DTQW}
The standard form of DTQW evolution on a two-state particle in 1D lattice is defined on a coin (c) and the position (p) Hilbert space ${\cal H} =  {\cal H}_c\otimes    {\cal     H}_p$. The basis states of ${\cal H}_c$ are the internal states of the particle, $|\downarrow \rangle = \begin{bmatrix} 1  \\ 0 \end{bmatrix}$ and $|\uparrow \rangle =\begin{bmatrix} 0  \\ 1 \end{bmatrix}$ and they are also the eigenstates of the Pauli matrix $\sigma_3= \begin{bmatrix} 1 & ~~0\\ 0 & -1 \end{bmatrix}$. The basis states of ${\cal H}_p$ is described in terms of $|\psi_z\rangle$, where $z \in {\mathbbm  I}$, the set of integers associated with the lattice sites. Each step evolution of 1D DTQW is described using a  quantum coin operation $B_{\sigma_3} (\theta)  \equiv \begin{bmatrix}\cos(\theta)   &   \sin(\theta)  \\ 
-\sin(\theta) &  \cos(\theta) 
\end{bmatrix}$\footnote{Though $B_{\sigma_3}(\xi, \theta, \zeta)\in {\rm SU(2)}$ describes the general form of evolution of each step of the DTQW, its main essence can be completely captured by assigning $\xi = \zeta = 0$.} 
which evolves the particle (coin) into the superposition of the basis states followed by the unitary shift operator 
$S_{\sigma_3}\equiv         \sum_z \left [  |\downarrow \rangle\langle
\downarrow|\otimes|\psi_{z-1}\rangle\langle   \psi_z|   +  | \uparrow \rangle\langle
\uparrow|\otimes |\psi_{z+1}\rangle\langle \psi_z| \right ]$,
which shift the state of the particle in superposition of the position space. Therefore, the effective operation for each step of the DTQW 
is written in the form:
\be
\label{Wop}
W_{\sigma_3} (\theta)\equiv S_{\sigma_3}[B_{\sigma_3}(\theta) \otimes  {\mathbbm 1}].
\ee
The state after the $t$ step evolution of the DTQW is given by,
$|\Psi_t\rangle=[W_{\sigma_3}(\theta)]^t|\Psi_{\rm in}\rangle$,
where $|\Psi_{\rm in}\rangle= \left ( \cos(\delta/2)| \downarrow \rangle + e^{i\eta}
\sin(\delta/2)|\uparrow \rangle \right )\otimes |\psi_{0}\rangle$,
is the initial state of the particle at position $z=0$. The coin parameter $\theta$ controls the variance of the probability distribution of the walk and $\theta \neq 0, \pi/2$ is required to spread the amplitude in superposition of position space.
\par
The net evolution of each step of the DTQW in the form of Eq.\,(\ref{Wop}), can also be generated by the time independent effective Hamiltonian $H_{\sigma_3}(\theta)$. That is,
$W_{\sigma_3}(\theta) \equiv e^{-iH_{\sigma_3}(\theta)\tau}$ where $\hbar = 1$ and $\tau$ is the time required to implement one step of the walk.
To obtain an expression for $H(\theta)$ we will expand Eq.\,(\ref{Wop}) and rewrite to the form:
\bea
\label{eq:combop}
W_{\sigma_3}(\theta)
=  \begin{bmatrix}
   \mbox{~~}\cos(\theta)e^{-i\hat{P}_Zl\tau}      &     &     \sin(\theta)e^{-i\hat{P}_Zl\tau}
  \\ - \sin(\theta)e^{+i\hat{P}_Zl\tau} & & \cos(\theta)e^{+i\hat{P}_Zl\tau} 
\end{bmatrix}  
\eea 
where $\hat{P}_Z$ is the momentum operator whose action on all position space in the $Z$ axis is local such that $e^{\pm i \hat{P}_Zl \tau} |\psi_z \rangle = |\psi_{z \pm l} \rangle$ \cite{ADZ93, Kem03}, $\tau$ is the time required to implement each step of walk length $l$ (lattice separation). Hereafter, we will consider the time required to implement the unit length of each step of walk to be one ($\tau = l = 1$). To obtain the Hamiltonian form, $W_{\sigma_3}(\theta) = e^{-i H_{\sigma_3}}(\theta)$ is written as  
\bea
\label{Ham}
-iH_{\sigma_3}(\theta) &=& \ln \begin{bmatrix}
   \mbox{~~}\cos(\theta) e^{-i\hat{P}_Z}     &     &    \sin(\theta)e^{-i\hat{P}_Z}
  \\ - \sin(\theta)e^{+i\hat{P}_Z}  & &  \cos(\theta)e^{+i\hat{P}_Z} 
\end{bmatrix}  \nonumber \\
&=&  \ln \,(A )  = V \,\ln(\,\lambda_{\sigma_3}\,)\, V^{-1}.
\eea
\be
\lambda_{\sigma_3} =\begin{bmatrix}
\lambda^-_Z & & 
0 \\
0 & & \lambda^+_Z \end{bmatrix}
\ee
is the diagonal matrix composed of the eigenvalues 
\be
\label{l1}
\lambda^{\mp}_Z = \cos(\theta) \cos(\hat{P}_Z) \mp \sqrt {\cos^2(\theta)\cos^2(\hat{P}_Z) - 1}
\ee
of matrix $A$ where $\lambda^{+}_Z \lambda^{-}_Z = \lambda^{-}_Z \lambda^{+}_Z = 1$. 
\bea
 V =\begin{bmatrix}
 \frac{\cos(\theta)e^{i \hat{P}_Z} - \lambda_{-}}{\sin(\theta) e^{i \hat{P}_Z}} & &  \frac{\cos(\theta)e^{i \hat{P}_Z} - \lambda_{+}}{\sin(\theta) e^{i \hat{P}_Z}} \\
1 & & 1 \end{bmatrix}
\eea
 and  
\bea
V^{-1} =\begin{bmatrix}
\frac{\sin(\theta) e^{i\hat{P}_Z}}{2 \sqrt{\cos^2(\theta)\cos^2(\hat{P}_Z) -1}} & & 
\frac{\lambda_{+} - \cos(\theta) e^{i\hat{P}_Z}} { 2 \sqrt{\cos^2(\theta)\cos^2(\hat{P}_Z) -1}} \\
\frac{-\sin(\theta)e^{+i\hat{P}_Z}}{2 \sqrt{\cos^2(\theta)\cos^2(\hat{P}_Z) -1}} & &
\frac{-\lambda_{-} + \cos(\theta) e^{i\hat{P}_Z}} { 2 \sqrt{\cos^2(\theta)\cos^2(\hat{P}_Z) -1} }\end{bmatrix} 
\eea
are the matrix composed of eigenvectors of $A$ and its inverse. By substituting these elements into Eq.\,(\ref{Ham}),
\begin{widetext}
\bea
\label{hamil2}
H_{\sigma_{3}}(\theta)& = &
\frac{ \ln \left (\frac{\lambda^{+}_Z}{\lambda^{-}_Z} \right )}{2 \sqrt{\cos^2(\theta)\cos^2(\hat{P}_Z) -1}}
\begin{bmatrix}
\cos(\theta) \sin(\hat{P}_Z)  & 
-\sin(\theta)[\sin(\hat{P}_Z)  + i \cos(\hat{P}_Z)]\\
\sin(\theta)[\sin(\hat{P}_Z)  - i \cos(\hat{P}_Z)] 
&
\cos(\theta) \sin(\hat{P}_Z) 
\end{bmatrix}
 \cdot \sigma_3
\eea
\end{widetext}
where $\sin(\hat{P}_Z) |\psi_{z} \rangle =   \frac{i}{2}(|\psi_{z-1} \rangle - |\psi_{z+1} \rangle)$ and $\cos(\hat{P}_Z) |\psi_{z} \rangle =   \frac{1}{2}(|\psi_{z-1} \rangle + |\psi_{z+1} \rangle)$.
\section{Two-state particle walk on different lattices}
\label{2sqw}
\subsection{Square Lattice and Cubic Lattice}
\label{square}
\begin{figure}[ht]
\bc 
\subfigure[]{\includegraphics[width=4.0cm]{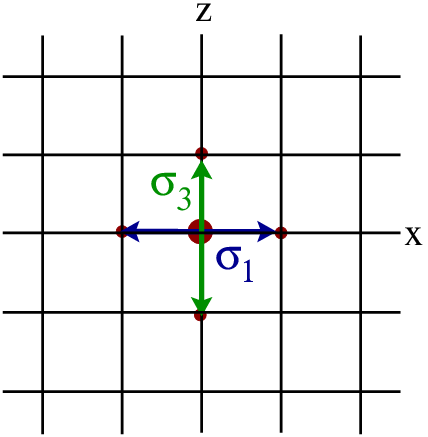}
\label{fig:1a}}
\subfigure[]{\includegraphics[width=4.0cm]{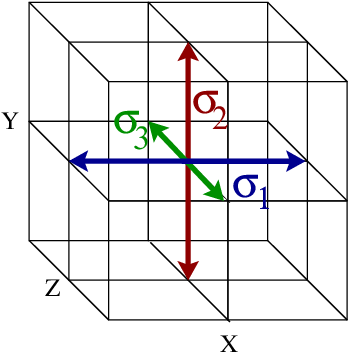}
\label{fig:1b}}
\caption{\footnotesize{(a) Square lattice with two direction of propagation in each $X$ and $Y$ axis. (b) Cubic lattice with two direction of propagation in each $X$, $Y$ and $Z$ axis. Evolution in $X$, $Y$ and $Z$ axis are quantized by the basis states of the Pauli operators $\sigma_{1}$, $\sigma_{2}$ and $\sigma_{3}$, respectively.}}
\ec 
\end{figure}

A simple 2D lattice structure is a square lattice with four direction of propagation and two quantization axis, $X$ and $Z$ [Fig.\,\ref{fig:1a}]. Similarly, a simple 3D lattice is a cubic lattice with six direction of propagation and three quantization axis, $X$, $Y$, and $Z$ [Fig.\,\ref{fig:1b}].
\par
Two-state particle DTQW on a square and cubic lattice can be realized by quantizing the evolution using different Pauli basis states as translational eigenstate for each  axis in the lattice structure.  
For a walk in 1D ($Z$ axis) we used basis states, $|\downarrow\rangle \equiv |+\rangle_{\sigma_{3}}$ and $|\uparrow\rangle \equiv |-\rangle_{\sigma_{3}}$ of Pauli operator $\sigma_{3}$ as translational state. Similarly, for $X$ and $Y$ axis we will use the basis states, 
\bea
 |+\rangle_{\sigma_{1}} = \frac{1}{\sqrt 2} \begin{bmatrix} 1 \\ 1 \end{bmatrix} ~;~ |-\rangle_{\sigma_{1}}= \frac{1}{\sqrt 2} \begin{bmatrix} 1 \\ -1 \end{bmatrix} \\
|+\rangle_{\sigma_{2}} = \frac{1}{\sqrt 2} \begin{bmatrix} 1 \\ i \end{bmatrix} ~;~ |-\rangle_{\sigma_{2}}= \frac{1}{\sqrt 2} \begin{bmatrix} 1 \\ -i \end{bmatrix}
\eea
of Pauli operators $\sigma_{1} = \begin{bmatrix} 0 & 1\\ 1 & 0 \end{bmatrix}$ and $\sigma_{2}= \begin{bmatrix} 0  & -i\\ i & ~0 \end{bmatrix}$ as translational states. The choice of a particular Pauli basis for particular axis is purely conventional and all the matrices hereafter are represented in the basis formed by the eigenvectors of $\sigma_{3}$.
The general form of the coin operation in any of the basis states $|\pm \rangle_{\sigma_{\alpha}}$ of the Pauli operators $\sigma_{\alpha}$ with $\alpha = 1, 2, 3$ will be  
\bea
\label{coinP}
B_{\sigma_{\alpha}}(\theta) = \cos (\theta) |+ \rangle_{\sigma_{_{\alpha}}} \langle + |
 + \sin (\theta) |+ \rangle_{\sigma_{\alpha}} \langle -| \nonumber  \\
 - \sin (\theta) |- \rangle_{\sigma_{\alpha}} \langle + |
+ \cos (\theta) |- \rangle_{\sigma_{\alpha}} \langle - |.
\eea
The shift operator for each axis of the square lattice ($Z$ and $X$) and the cubic lattice ($Z$, $X$, $Y$) will be
\begin{subequations}
\begin{eqnarray}
\label{shiftS}
S_{\sigma_3}^{sq} \equiv \sum_{x, z}  [ |+ \rangle_{\sigma_{3}}\langle
+|\otimes|\psi_{x, z-1}\rangle\langle   \psi_{x, z}|  \nonumber \\ 
+    | - \rangle_{\sigma_{3}}\langle
-|\otimes |\psi_{x, z+1}\rangle\langle \psi_{x, z}|],  \\
S_{\sigma_1}^{sq} \equiv \sum_{x, z}  [ |+ \rangle_{\sigma_{1}}\langle
+|\otimes|\psi_{x-1, z}\rangle\langle   \psi_{x, z}|  \nonumber \\
+    | - \rangle_{\sigma_{1}}\langle  
-|\otimes |\psi_{x+1, z}\rangle\langle \psi_{x, z}|], \\
S_{\sigma_3}^{cub} \equiv \sum_{x,y,z}  [ |+ \rangle_{\sigma_{3}}\langle
+|\otimes|\psi_{x, y,z-1}\rangle\langle   \psi_{x, y, z}|  \nonumber \\
+    | - \rangle_{\sigma_{3}}\langle
-|\otimes |\psi_{x,y, z+1}\rangle\langle \psi_{x, y, z}|],  \\
S_{\sigma_1}^{cub} \equiv \sum_{x, y, z}  [ |+ \rangle_{\sigma_{1}}\langle
+|\otimes|\psi_{x-1, y, z}\rangle\langle   \psi_{x,y, z}|  \nonumber \\
+    | - \rangle_{\sigma_{1}}\langle  
-|\otimes |\psi_{x+1,y, z}\rangle\langle \psi_{x,y, z}|], \\
S_{\sigma_2}^{cub} \equiv \sum_{x, y, z}  [ |+ \rangle_{\sigma_{2}}\langle
+|\otimes|\psi_{x, y-1, z}\rangle\langle   \psi_{x,y, z}|  \nonumber \\
+    | - \rangle_{\sigma_{1}}\langle  
-|\otimes |\psi_{x,y+1, z}\rangle\langle \psi_{x,y, z}|],
\end{eqnarray}
\end{subequations}
where position state $|\psi_{x, y, z}\rangle = |\psi_{x}\rangle \otimes |\psi_{y}\rangle \otimes |\psi_{z}\rangle$. One complete step of the DTQW on a square and cubic lattice using two-state particle  composes of the evolution in one axis followed by the evolution in the other, that is,
\begin{subequations}
\bea
\label{w2d}
W^{sq}(\theta) = W_{\sigma_1}^{sq} (\theta) W_{\sigma_3}^{sq} (\theta),\\
\label{w3d}
W^{cub}(\theta) = W_{\sigma_2}^{cub} (\theta)W_{\sigma_1}^{cub} (\theta) W_{\sigma_3}^{cub} (\theta),
\eea
\end{subequations}
where $W_{\sigma_\alpha}^{sq} (\theta)= S_{\sigma_\alpha}^{sq} [ B_{\sigma_{\alpha}}(\theta) \otimes  {\mathbbm 1}_X\otimes  {\mathbbm 1}_Z]$ and $W_{\sigma_\alpha}^{cub} (\theta)=S_{\sigma_\alpha}^{cub} [ B_{\sigma_{\alpha}}(\theta) \otimes  {\mathbbm 1}_X\otimes  {\mathbbm 1}_Y \otimes  {\mathbbm 1}_Z]$, respectively.  Equivalent form of the evolution operators are,
\begin{subequations}
\begin{eqnarray}
W_{\sigma_3}^{sq} (\theta) \equiv{\mathbbm 1}_X \otimes W_{\sigma_3}(\theta) = {\mathbbm 1}_X \otimes  e^{-iH_{\sigma_3}(\theta)} ~\\
W_{\sigma_1}^{sq} (\theta)\equiv W_{\sigma_1} (\theta)\otimes {\mathbbm 1}_Z \equiv e^{-iH_{\sigma_1}(\theta)} \otimes  {\mathbbm 1}_Z~\\
W_{\sigma_3}^{cub} (\theta) \equiv  {\mathbbm 1}_X\otimes  {\mathbbm 1}_Y \otimes W_{\sigma_{3}}(\theta)
= {\mathbbm 1}_X\otimes  {\mathbbm 1}_Y \otimes  e^{-iH_{\sigma_3}(\theta)}~\\
W_{\sigma_1}^{cub} (\theta) \equiv W_{\sigma_{1}}(\theta) \otimes {\mathbbm 1}_Y \otimes  {\mathbbm 1}_Z  
=  e^{-iH_{\sigma_1}(\theta)}\otimes  {\mathbbm 1}_Y \otimes  {\mathbbm 1}_Z ~\\
W_{\sigma_2}^{cub} (\theta) \equiv  {\mathbbm 1}_X \otimes W_{\sigma_{2}}(\theta) \otimes  {\mathbbm 1}_Z  
=    {\mathbbm 1}_X \otimes e^{-iH_{\sigma_2}(\theta)} \otimes  {\mathbbm 1}_Z~. 
\end{eqnarray}
\end{subequations}
The operator $W_{\sigma_3}(\theta)$ and $H_{\sigma_3}(\theta)$ are  given by Eq.\,(\ref{eq:combop}) and Eq.\,(\ref{hamil2}). Similarly, the operator for $X$ and $Y$ axis,
\begin{subequations}
\bea
&W_{\sigma_1} (\theta) = S_{\sigma_1} [ B_{\sigma_{1}}(\theta) \otimes  {\mathbbm 1}] \equiv e^{-iH_{\sigma_1}(\theta)} \\
\label{eq:combop33}
 &= \frac{e^{-i\hat{P}_X}}{2} \begin{bmatrix}
   \cos(\theta)+ \sin(\theta)     &   \cos(\theta)-\sin(\theta)\\ 
  \cos(\theta)+\sin(\theta) &   \cos(\theta)-\sin(\theta) 
\end{bmatrix} + \nonumber \\
&  \frac{e^{+i\hat{P}_X}}{2} \begin{bmatrix}
 \cos(\theta)-\sin(\theta)  &  -\cos(\theta)-\sin(\theta)\\ 
 -\cos(\theta)+\sin(\theta) & \cos(\theta)+\sin(\theta)
\end{bmatrix};\\
& W_{\sigma_2} (\theta) =  S_{\sigma_2} [ B_{\sigma_{2}}(\theta) \otimes  {\mathbbm 1}] \equiv e^{-iH_{\sigma_2}(\theta)} \\
\label{eq:combop4}
 &=\frac{e^{-i\hat{P}_Y}}{2}\begin{bmatrix}
   \cos(\theta)+ \sin(\theta)         &    -i\cos(\theta)+i\sin(\theta)
  \\ i\cos(\theta)+i\sin(\theta)  &  \cos(\theta)-\sin(\theta) 
\end{bmatrix} \nonumber \\
  &+\frac{e^{i\hat{P}_Y}}{2} \begin{bmatrix}
  \cos(\theta)-\sin(\theta)           &  i\cos(\theta)+i\sin(\theta)
  \\ -i\cos(\theta)+\sin(\theta) & \cos(\theta)+\sin(\theta) 
\end{bmatrix},
\eea
\end{subequations}
where $\hat{P}_X$ and $\hat{P}_Y$ are the momentum operator on $X$ and $Y$ axis.
The eigenvalues, $\lambda^\mp_X$ of $W_{\sigma_1}(\theta)$ and $\lambda^\mp_Y$ of $W_{\sigma_2}(\theta)$ are same as the eigenvalues $\lambda^\mp_Z$ with only a replacement of $\hat{P}_X$ and $\hat{P}_Y$ in place of $\hat{P}_Z$ in Eq.\,(\ref{l1}). Therefore, using the eigenvalues, eigenvectors and inverse of eigenvector, the Hamiltonian form for the evolution in $X$ and $Y$ axis are 
\begin{widetext}
\begin{subequations}
\bea
\label{hamil2b}
H_{\sigma_{1}}(\theta)= 
\frac{\ln \left (\frac{\lambda^{+}_X}{\lambda^{-}_X} \right )}{2 \sqrt{\cos^2(\theta)\cos^2(\hat{P}_X) -1}}
\begin{bmatrix}
\cos(\theta)\sin(\hat{P}_X)  - i \cos(\hat{P}_X)\sin(\theta) \,  & 
\sin(\theta) \sin(\hat{P}_X)\, \\
-\sin(\theta) \sin(\hat{P}_X) \, 
&
\cos(\theta)\sin(\hat{P}_X)  + i \cos(\hat{P}_X)\sin(\theta)
\end{bmatrix} \cdot \sigma_1 \\
\label{hamil2d}
H_{\sigma_{2}}(\theta) = 
\frac{\ln \left (\frac{\lambda^{+}_Y}{\lambda^{-}_Y} \right )}{2 \sqrt{\cos^2(\theta)\cos^2(\hat{P}_Y) -1}}
\begin{bmatrix}
\cos(\theta)\sin(\hat{P}_Y)-i \cos(\hat{P}_Y)\sin(\theta) & 
-i\sin(\theta) \sin(\hat{P}_Y)\\
-i\sin(\hat{P}_Y) \sin(\theta) 
&
\cos(\theta)\sin(\hat{P}_Y)+i\cos(\hat{P}_Y)\sin(\theta)  
\end{bmatrix} \cdot \sigma_2
\eea
\end{subequations}
\end{widetext}
When $\theta = 0$, that is, in absence of a coin operation $\lambda^{\mp}_Z = e^{\mp i\hat{P}_Z}$,  $\lambda^{\mp}_X = e^{\mp i\hat{P}_X}$, and  $\lambda^{\mp}_Y = e^{\mp i\hat{P}_Y}$. This reduces the Hamiltonian form to, $H_{\sigma_{3}}(0) = 
\begin{bmatrix}
\hat{P}_Z  &   0 \\
0 &  \hat{P}_Z  
\end{bmatrix} \cdot \sigma_3$, $H_{\sigma_{1}}(0) = 
\begin{bmatrix}
\hat{P}_X   & 0 \\
0 &  \hat{P}_X  
\end{bmatrix} \cdot \sigma_1$, and $H_{\sigma_{2}}(0) = 
\begin{bmatrix}
\hat{P}_Y   & 0 \\
0 &  \hat{P}_Y  
\end{bmatrix} \cdot \sigma_2$, respectively.
\par
If the initial state of the particle on a square lattice $|C \rangle=\frac{1}{\sqrt{2}}[|\downarrow\rangle + i |\uparrow \rangle]$, 
$|\Psi_{\rm in} \rangle = \frac{1}{\sqrt 2}\begin{bmatrix}
[|\psi_{x_0}\rangle \otimes  |\psi_{z_0}\rangle] \\
i[|\psi_{x_0}\rangle \otimes  |\psi_{z_0}\rangle]
\end{bmatrix}$ and the state after $t$ step [$(W^{sq}(\theta))^t$],
\bea 
|\Psi_t\rangle = \sum_{x =-t}^t \sum_{z =-t}^t \Big [ \alpha^{(1)}_{(x, z, t)}|\downarrow
  \rangle  + \alpha^{(2)}_{(x, z, t)}|\uparrow \rangle \Big ] 
\otimes|\psi_{x,  z}\rangle.
\label{eq:lr2}
\eea
When $\theta =0$,  $\alpha^{(1)}_{(x, y,t)}$ and $\alpha^{(2)}_{(x, y, t)}$ are given by the
coupled iterative relations
\begin{subequations}
\label{eq:iter}
\begin{eqnarray}
\alpha^{(1)}_{(x, y,t)} = \frac{1}{2}\Big [  \alpha^{(1)}_{(x+1, y+1,t-1)}  + \alpha^{(1)}_{(x+1, y-1,t-1)} \nonumber \\
+ \alpha^{(2)}_{(x-1, y+1,t-1)} -  \alpha^{(2)}_{(x-1, y-1,t-1)}\Big ]  \\
\alpha^{(2)}_{(x, y,t)} = \frac{1}{2}\Big [  \alpha^{(1)}_{(x+1, y+1,t-1)} -  \alpha^{(1)}_{(x+1, y-1,t-1)} \nonumber \\
+ \alpha^{(2)}_{(x-1, y+1,t-1)} +  \alpha^{(2)}_{(x-1, y-1,t-1)}\Big ].
\end{eqnarray}
\end{subequations}
\begin{figure}[ht]
\bc
\subfigure[]{\includegraphics[width=63mm]{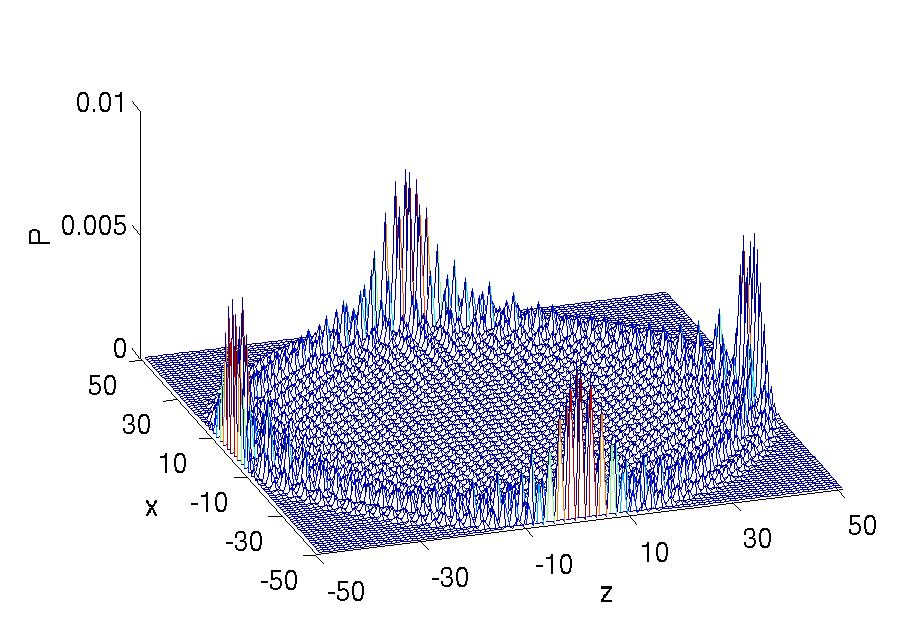}
\label{fig:3a}}
\subfigure[]{\includegraphics[width=63mm]{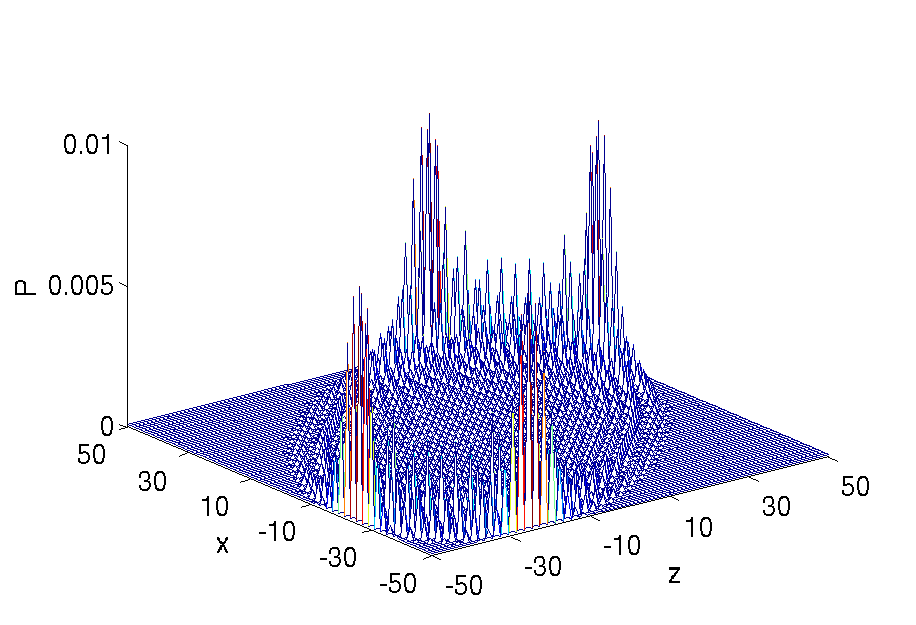}\label{fig:3b}}
\caption{\footnotesize{Probability distribution of a 50 step two-state particle DTQW with the initial state $|\Psi_{\rm in} \rangle = \frac{1}{\sqrt 2}[|\downarrow \rangle + i|\uparrow \rangle]\otimes|\psi_{x_0}\rangle \otimes  |\psi_{z_0}\rangle$ on a square lattice using basis state of different Pauli operator for each axis ($\sigma_1$ for $X$ axis and $\sigma_3$ for $Z$ axis). (a) The distribution is after the evolution without the coin operation in both the axis ($\theta =0$) and same distribution is obtained for Grover walk. (b) The distribution is after the evolution with the coin operation $\theta = \pi/12$ in both the axis.}}
\ec
\end{figure}
In Fig.\,\ref{fig:3a}, the probability distribution of the 50 step DTQW on a square lattice using the Pauli basis scheme without the coin operation ($\theta = 0$) is shown. This probability distribution obtained is identical to the ones reported for Grover walk on a four-state particle \cite{TFM03} and for the alternative walk on a two-state particle with initial state, $\frac{1}{\sqrt{2}}[|\downarrow \rangle + i |\uparrow \rangle]$ using Hadamard operator as the coin operation \cite{FGB11}.  Each step of Grover walk is on a 2D is realized using the Grover diffusion operator 
$G = \frac{1}{2}\begin{bmatrix}-1 & ~~1 & ~~1 & ~~1  \\
  ~~1 & -1 & ~~1 & ~~1  \\
  ~~1 & ~~1 & -1 & ~~1  \\
  ~~1 & ~~1 & ~~1 & -1
\end{bmatrix}$ as coin operation Followed by $
  S^G \equiv \sum_{x, z} \Big[|\downarrow \rangle\langle \downarrow|\otimes|
                          \psi_{x-1, z-1}\rangle\langle \psi_{x, z}|   
                        +     |\uparrow \rangle\langle \uparrow|\otimes |
                          \psi_{x-1, z+1}\rangle\langle \psi_{x, z}| 
                        +     |\leftarrow \rangle\langle \leftarrow|\otimes|
                          \psi_{x+1, z-1}\rangle\langle   \psi_{x, z}| 
                        +     |\rightarrow \rangle\langle \rightarrow|\otimes |
                          \psi_{x+1, z+1}\rangle\langle   \psi_{x, z}| \Big]$ on a particle in a specific initial state, $|\Psi_{\rm in}^{G}\rangle=\frac{1}{2}[|\downarrow \rangle -|\uparrow \rangle -|\leftarrow \rangle +|\rightarrow \rangle]$. 
The state after $t$ step of the Grover walk [$S^G(G \otimes  {\mathbbm 1})^t]$,
\bea 
|\Psi^G_t\rangle = \sum_{x =-t}^t \sum_{z =-t}^t \Big [ \beta^{(1)}_{(x, z, t)}|\downarrow
  \rangle  + \beta^{(2)}_{(x, z, t)}|\uparrow \rangle +  \beta^{(3)}_{(x, z, t)}|\leftarrow \rangle \nonumber \\
 + \beta^{(4)}_{(x, z, t)}|\rightarrow \rangle \Big ] 
\otimes|\psi_{x,  z}\rangle,
\label{eq:lr2}
\eea
where $\beta(x, y,t)$'s  are given by the
quadrupled iterative relation coupling the $X$ and $Z$ axis
\begin{subequations}
  \label{eq:4s_iter}
  \begin{eqnarray}
   \beta^{(1)}(x, z, t) &= \frac{1}{2} \Big[-\beta^{(1)}_{(x+1, z+1,t-1)} + \beta^{(2)}_{(x+1, z+1,t-1)} \nonumber \\
    & + \beta^{(3)}_{(x+1, z+1,t-1)} +  \beta^{(4)}_{(x+1, z+1,t-1)}\Big ] \\
   \beta^{(2)}(x, z, t) &= \frac{1}{2} \Big[\beta^{(1)}_{(x+1, z-1,t-1)} -  \beta^{(2)}_{(x+1, z-1,t-1)}  \nonumber \\
    &+ \beta^{(3)}_{(x-1, z-1,t-1)} +  \beta^{(4)}_{(x+1, z-1,t-1)} \Big] \\
    \beta^{(3)}(x, z, t) &= \frac{1}{2} \Big[\beta^{(1)}_{(x-1, z+1,t-1)} + \beta^{(2)}_{(x-1, z+1,t-1)} \nonumber \\
    & - \beta^{(3)}_{(x-1, z+1,t-1)}  + \beta^{(4)}_{(x-1, z+1,t-1)}\Big ]\\
    \beta^{(4)}(x, z, t) &= \frac{1}{2} \Big [\beta^{(1)}_{(x-1, z-1,t-1)} +  \beta^{(2)}_{(x-1, z-1,t-1)} \nonumber \\
    & + \beta^{(3)}_{(x-1, z-1,t-1)} -  \beta^{(4)}_{(x-1, z-1,t-1)} \Big ].
  \end{eqnarray}
\end{subequations}
From Eqs.\,(\ref{eq:iter}) and Eqs.\,(\ref{eq:4s_iter}) we can note that for both, two-state walk and the Grover walk, the amplitude at any position $(x, z)$ for a given time $t$ is dependent on the amplitude at the four diagonally opposite sites at time $t-1$. Therefore, starting from a specific initial state of two-state and four-state particle as discussed in this section, these amplitudes returns the same probability distribution. 
\par
Unlike the Grover walk which is very specific to the initial state and the coin operation, probability distribution with the two-state walk using the Pauli basis can be controlled by introducing the coin operation ($\theta \neq 0$) and/or using different initial state of the particle. In Fig.\,\ref{fig:3b} the probability distribution of the 50 step DTQW with coin operation, $\theta = \pi/12$ is show to squeeze the distribution towards the diagonal of the square lattice.
\par
The basis states of $\sigma_3$ can be written as a superposition of a basis states of the $\sigma_1$. Therefore, even in absence of a coin operation, $[W^{sq}_{\sigma_{1}}(0)W^{sq}_{\sigma_{3}}(0)]^t$ evolves the particle in superposition of position space and implement a DTQW on a square lattice. Similarly,  for a two-state walk on a cubic lattice using different basis states, due to the relationship between the basis states of the Pauli operators,  the particle evolves in superposition of position space even in absence of coin operation.  However, the coin operation can be effectively used to control the dynamics and the probability distribution of the walk. 

\subsection{Triangular lattice}
\begin{figure}[ht]
\bc
\includegraphics[width=7.6cm]{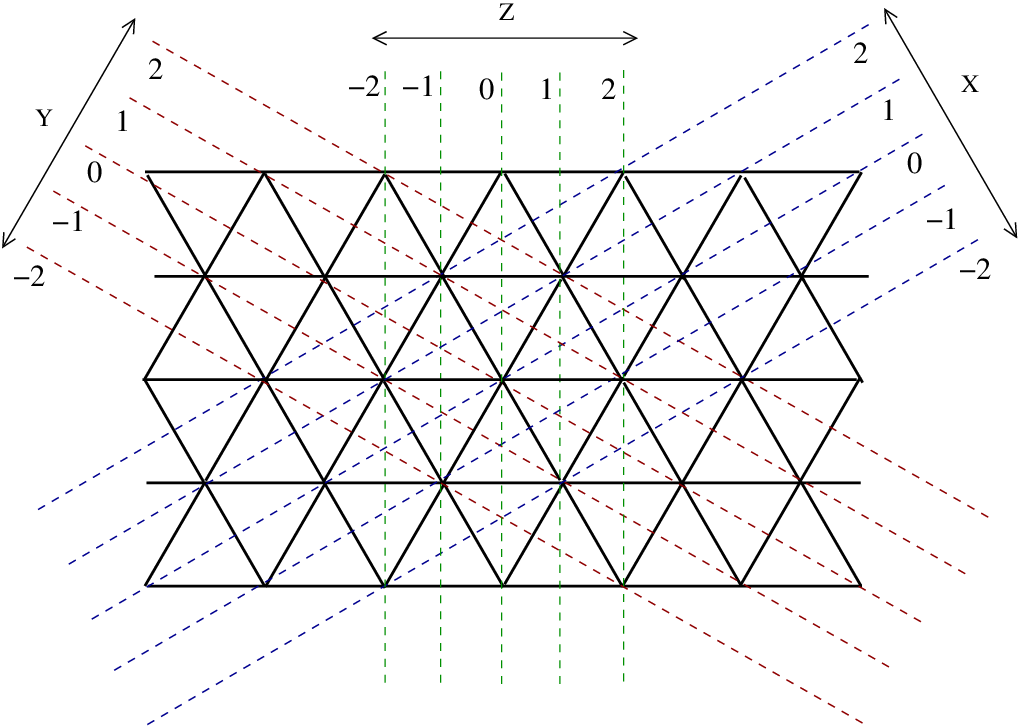}
\caption{\footnotesize{ Triangular lattice structure with labeling of lattice position in all the three axis, $X$, $Y$, and $Z$ of propagation quantized by the basis states of the Pauli operators $\sigma_{1}$, $\sigma_{2}$ and $\sigma_{3}$, respectively.}}
\label{fig:4}
\ec
\end{figure}

The triangular lattice structures shown in Fig.~\ref{fig:4} also has a three axis of propagation, $X, Y$, and $Z$. Therefore, the walk can be quantized using the eigenstates, $|+\rangle_{\sigma_{\alpha}}$ and $|-\rangle_{\sigma_{\alpha}}$ of the Pauli operators $\sigma_{\alpha}$ where $\alpha  =$ 1, 2, and 3, as translational basis states. 

\begin{figure}[ht]
\subfigure[]{
\includegraphics[width=5.5cm]{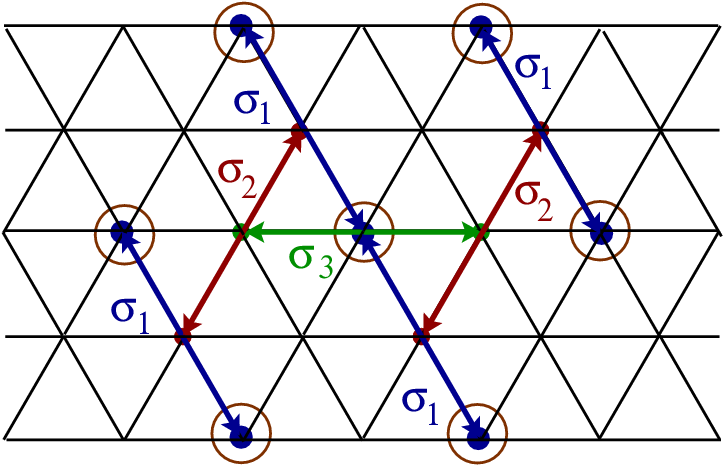}
\label{fig:5}}
\subfigure[]{\includegraphics[width=6.2cm]{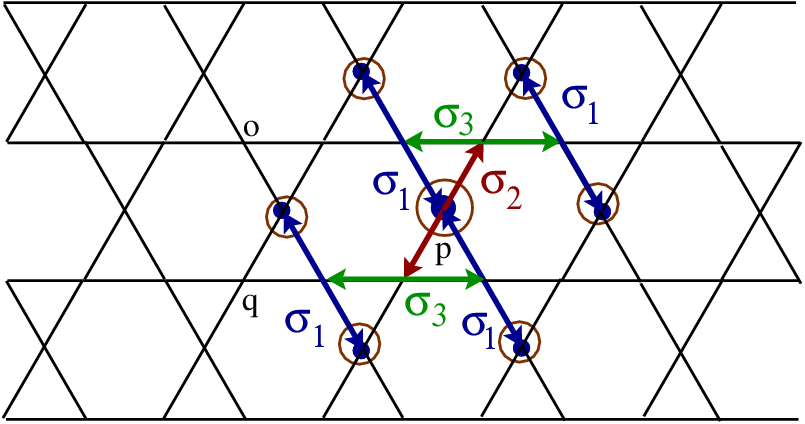}
\label{fig:6}}
\caption{\footnotesize{ Scheme for evolution of DTQW on: (a) The triangular lattice, starting from the middle, the arrow marks show the shift in position space during one step of DTQW evolution (evolution in $Z$ axis followed by the evolution in the $Y$ and $X$ axis). (b) Kagome lattice structure with two axis of propagation at each lattice site. From lattice sites {\bf o}, {\bf p}, and {\bf q}, we can see that they are associated with different combination of quantization axis.  Starting from position {\bf p}, the arrow marks show the shift in position space during one step of DTQW evolution. The final positions are encircled.}}
\end{figure}
Unlike the cubic lattice the three axis for propagation in triangular lattice are not orthogonal to each other and hence, the evolution in one axis alters the evolution in other two axis as well. 
Therefore, the shift operator has to be defined according to the lattice structure. Shift operator in each axis can be defined such that the unit shift in the main axis is accompanied by half of the unit shift in the other two axis. In Fig.\,\ref{fig:4}, the labeling of the position is shown and for convenience we choose the two unit position shift in the main axis and one unit position shift in the other two axis. Therefore, the shift operation for $\sigma_3$ basis states is,
\bea
S_{\sigma_3} \equiv \sum_{x, y, z}  |\downarrow \rangle\langle
\downarrow |\otimes|\psi(x+1, y-1, z-2)\rangle\langle   \psi(x, y, z)| \nonumber \\ 
+   \sum_{x, y, z} | \uparrow \rangle\langle
\uparrow |\otimes |\psi(x-1, y+1, z+2)\rangle\langle \psi(x, y, z)|.
\eea
and the effective evolution operator
\bea
\label{eq:comboptri}
W^{\prime}_{\sigma_3}(\theta) = \begin{bmatrix}
   \mbox{~~}\cos(\theta) e^{-i\hat{P}_3}     &     &    \sin(\theta)e^{-i\hat{P}_3}
  \\ - \sin(\theta)e^{+i\hat{P}_3}  & &  \cos(\theta)e^{+i\hat{P}_3} 
\end{bmatrix}, 
\eea
where $\hat{P}_3 = -\hat{P}_{X} \otimes \hat{P}_{Y} \otimes 2\hat{P}_{Z}$. Therefore, the Hamiltonian $H^{\prime}_{\sigma_{3}}(\theta)$ will be in the same form of Eq. (\ref{hamil2}) with a replacement of $\hat{P}_{Z}$ by $\hat{P}_3$. Similarly, the evolution operator $W^{\prime}_{\sigma_1}(\theta)$ and $W^{\prime}_{\sigma_2}(\theta)$, and the Hamiltonian form in the $X$ and $Y$ axis, $H^{\prime}_{\sigma_{1}}(\theta)$ and $H^{\prime}_{\sigma_{2}}(\theta)$, will be in the same form of Eqs. (~\ref{eq:combop33}),~(\ref{eq:combop4}) and Eqs. (~\ref{hamil2b}),~(\ref{hamil2d}), respectively with the replacement of $\hat{P}_{X}$ by $\hat{P}_1 =  2\hat{P}_{X} \otimes \hat{P}_{Y} \otimes -\hat{P}_{Z}$ and  $\hat{P}_{Y}$ by $\hat{P}_2 =  -\hat{P}_{X} \otimes 2\hat{P}_{Y} \otimes \hat{P}_{Z}$.
\begin{figure}[ht]
\bc
\subfigure[]{\includegraphics[width=70.5mm]{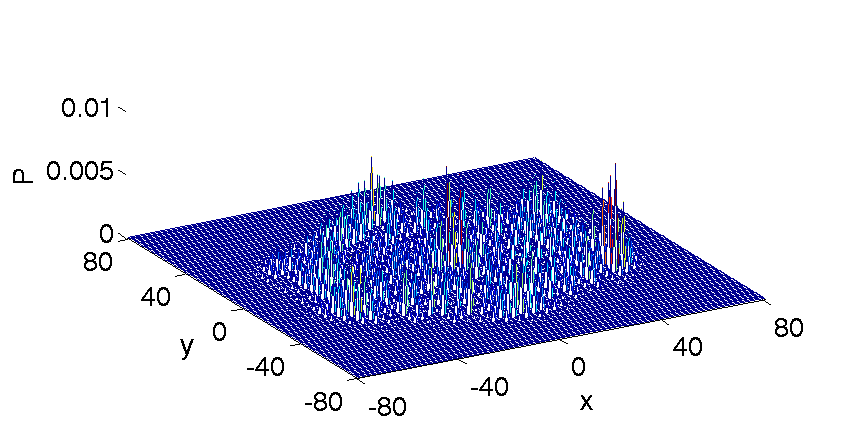}
\label{fig:5a}}
\hskip -0.17in
\subfigure[]{\includegraphics[width=70.5mm]{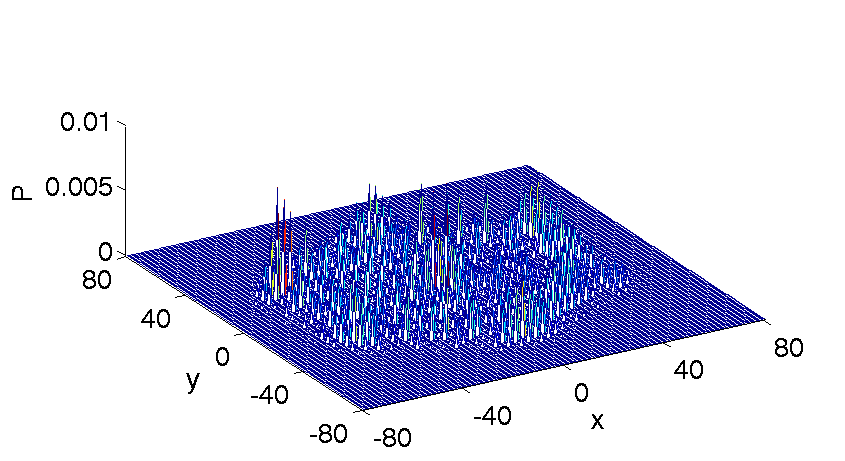}
\label{fig:5b}}
\vskip -0.25cm
\subfigure[]{\includegraphics[width=70mm]{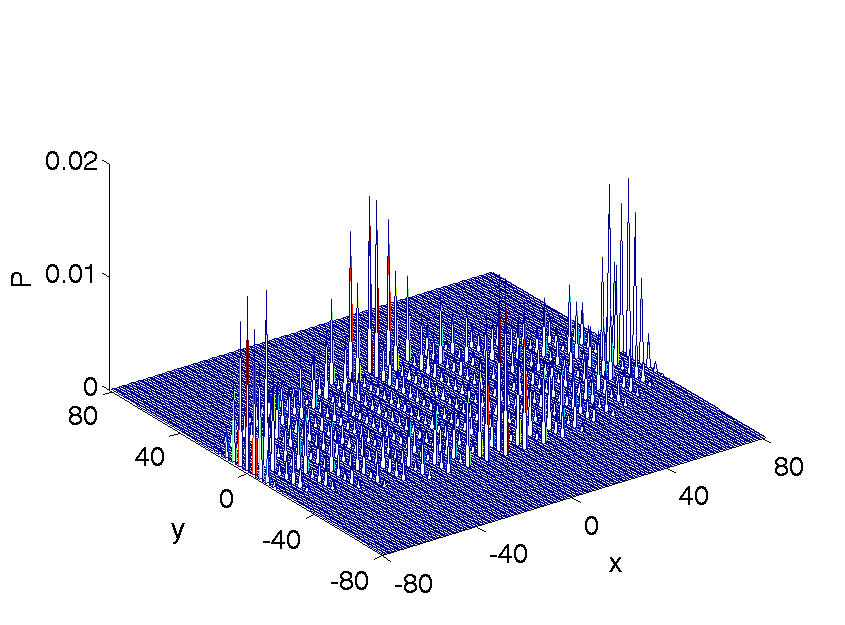}
\label{fig:5c}}
\caption{\footnotesize{Probability distribution of 20 step two-state particle DTQW on triangular lattice. 
(a) The initial state is $|\Psi_{\rm in} \rangle = |\downarrow \rangle \otimes|\psi_{x_0}\rangle \otimes |\psi_{y_0}\rangle \otimes |\psi_{z_0}\rangle$ and the walk is evolved without a coin operation. (b) The initial state is $|\Psi_{\rm in} \rangle = |\uparrow \rangle \otimes|\psi_{x_0}\rangle \otimes |\psi_{y_0}\rangle \otimes |\psi_{z_0}\rangle$ and the walk is evolved without a coin operation. (c) The initial state is $|\Psi_{\rm in} \rangle = |\downarrow \rangle \otimes|\psi_{x_0}\rangle \otimes |\psi_{y_0}\rangle \otimes |\psi_{z_0}\rangle$ and the walk is evolved with one coin operation [$H^{\prime}_{\sigma_{1}}(0)+ H^{\prime}_{\sigma_{2}}(\pi/4)+ H^{\prime}_{\sigma_{3}}(0)$] to obtain a symmetric probability distribution in position space.}}
\ec
\end{figure}
Each Hamiltonian, $H^{\prime}_{\sigma_{1}}(\theta)$, $H^{\prime}_{\sigma_{2}}(\theta)$, and $H^{\prime}_{\sigma_{3}}(\theta)$ evolve the state enabling the interaction between the three  quantization axis. 
Therefore, each step of DTQW can be realized by the evolution using one Pauli basis followed by the other, $W^{tri}(\theta) = W^{\prime}_{\sigma_1}(\theta)W^{\prime}_{\sigma_2}(\theta)W^{\prime}_{\sigma_3}(\theta)$ as shown in Fig.~\ref{fig:5}. We should note that the three effective Hamiltonian for the operators, $H^{\prime}_{\sigma_{1}}(\theta)$, $H^{\prime}_{\sigma_{2}}(\theta)$, and $H^{\prime}_{\sigma_{3}}(\theta)$ together commute and complete Hamiltonian for each step of DTQW on triangular lattice can be written as
\bea
\label{hamil55}
H^{\prime}(\theta) &=& H^{\prime}_{\sigma_{1}}(\theta) +  H^{\prime}_{\sigma_{2}}(\theta) + H^{\prime}_{\sigma_{3}}(\theta).
\eea
The choice of the order of the basis in which the particle is evolved is purely conventional for the triangular lattice. Even when $\theta =0$, due to the interplay between different Pauli basis for translation in each axis,  a two-state particle evolve in superposition of position space resulting in a diffused probability distribution. However, a coin operation with different $\theta$ for each axis can be extensively used for the evolution to get addition freedom to control the evolution and obtained the desired probability distribution. In Fig.\,\ref{fig:5a} and \,\ref{fig:5b}, we show the probability distribution of a 20 step DTQW without a coin operation [$H^{\prime}(0) = H^{\prime}_{\sigma_{1}}(0) +  H^{\prime}_{\sigma_{2}}(0) + H^{\prime}_{\sigma_{3}}(0)$] on a two-state particle initially in state $|\downarrow \rangle$ and $|\uparrow \rangle$, respectively.  We can see that the probability distribution in Fig.\,\ref{fig:5a} and Fig.\,\ref{fig:5b} are not symmetric distribution in position space but are symmetric to each other.  In Fig.\,\ref{fig:5c} we show the symmetric probability distribution obtained by introducing a coin operation with $\theta = \pi/4$ for only one operation during each step evolution  [$H^{\prime}_{\sigma_{1}}(0) +  H^{\prime}_{\sigma_{2}}(\pi/4) + H^{\prime}_{\sigma_{3}}(0)$].

\subsection{Kagome Lattice}

Kagome lattice structure can also be labeled the same way as the triangular lattice. The evolution operator and its Hamiltonian form in each basis ($H^{\prime}_{\sigma_{1}}(\theta)$, $H^{\prime}_{\sigma_{2}}(\theta)$, and $H^{\prime}_{\sigma_{3}}(\theta)$)  will be in the same form as presented for triangular lattice. But, unlike triangular lattice which has three quantization axis at each lattice site,  kagome lattice shown in  Fig. \ref{fig:6} has only two quantization axis with four direction of propagation for  the walk at each lattice site. The two quantization axis at each lattice site is not the same for all lattice sites. In Fig. \ref{fig:6}, lattice sites {\bf o}, {\bf p}, and {\bf q} have axis $X$ and $Z$ ($\sigma_1$ and $\sigma_3$), $X$ and $Y$ ($\sigma_1$ and $\sigma_2$), and $Y$ and $Z$ ($\sigma_2$ and $\sigma_3$) as quantization axis, respectively. Therefore, to implement each step of DTQW in kagome lattice certain simple order for using evolution operators in different axis has to be followed. For example, if the initial position is {\bf p} (as marked in Fig.~\ref{fig:6}), each step of DTQW can be realized by $W^{kag}(\theta) = W^{\prime}_{\sigma_1}(\theta)W^{\prime}_{\sigma_3}(\theta)W^{\prime}_{\sigma_2}(\theta)$ and the effective Hamiltonian form for each step of DTQW will be 
$H^{\prime}(\theta) = H^{\prime}_{\sigma_{1}}(\theta) +  H^{\prime}_{\sigma_{3}}(\theta) + H^{\prime}_{\sigma_{2}}(\theta)$.
The main consideration in choosing the first axis is, not to pick the evolution in the axis which is nonexistent in that initial position.

\section{Conclusion} 
\label{conc}

In this paper we presented a new scheme for the evolution of DTQW on 2D and 3D lattices using a two-state particle. Our scheme used different Pauli basis states as translational eigestate in different axis and showed that the coin operation is not a necessary requirement to implement a walk on 2D and 3D systems but can be used as an additional degree of freedom to control the dynamics. We also discussed the Hamiltonian form of evolution for the walk which can serve as a general framework to simulate, control, and study the dynamics in different physical systems. The Pauli basis states for translation is commonly used to describe the dynamics in various physical systems, in particular, in quantum optics and optical lattice \cite{DDL03}. Therefore, use of Pauli basis state for translation without the use of coin operation and the Hamiltonian form can serve as a frame work for experimental implementation of DTQW with a minimum resource in various 2D and 3D physical structures.  Our scheme for evolution  on square, cubic, triangular, and kagome lattice can be straight away extended to other 2D and 3D Bravais lattice and the extension to other higher dimensions is also possible by permuting the three Pauli basis states for each translational axis. This description of dynamics in Hamiltonian form helps to further explore topological phase, establish connection between physical process in nature which are generally not 1D and does not involve larger internal (more than two) dimension of the particle. 

\end{document}